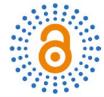

# Genoautotomy (Genome "Self-Injury") in Eukaryotic Cells: A Cellular Defence Response to Genotoxic Stress

Gao-De Li

Chinese Acupuncture Clinic, Liverpool, UK
Email: gaode_li@yahoo.co.uk

## Abstract

**This paper proposes that eukaryotic cells, under severe genotoxic stress, can commit genoautotomy (genome "self-injury") that involves cutting and releasing single-stranded DNA (ssDNA) fragments from double-stranded DNA and leaving ssDNA gaps in the genome. The ssDNA gaps could be easily and precisely repaired later. The released ssDNA fragments may play some role in the regulation of cell cycle progression. Taken together, genoautotomy causes limited nonlethal DNA damage, but prevents the whole genome from lethal damage, and thus should be deemed as a eukaryotic cellular defence response to genotoxic stress.**

## Keywords

## 1. Introduction

Both apoptosis ("self-killing") and autophagy ("self-eating") are cytoprotective responses to cellular stress, and work coordinately or independently to decide which cells live or die [1] [2]. Recently, we published a paper in which a new term, genome "self-injury" (autotomy), was first mentioned [3]. Originally, the term only referred to endonuclease dependent transcript cutout which is a hypothesis about the mechanism of DNA to DNA transcription [3] [4]. However, after rethinking the role of genome "self-injury" in cell survival, we realised that genome "self-injury" is such a big concept that endonuclease dependent transcript cutout is only one type of genome "self-injury". Therefore, in this paper, we would like to present a new thought on the role of genome "self-injury" in eukaryotic cells.

## 2. Genoautotomy: A Cellular Defence Response to Genotoxic Stress

Genoautotomy is a new term created for genome "self-injury", which implies defensive "self-injury" because autotomy, also called self-amputation, is defined as the shedding of a body part, which helps to prevent the whole animal from being compromised in response to external stimuli [5] [6]. We propose that like apoptosis and autophagy, genoautotomy could be another genetically programmed cellular defence response to genotoxic stress in eukaryotic cells, which is triggered by severe genotoxic stress caused by certain endogenous genotoxic factors (genotoxic subatances produced by cellular metabolism or obstacles encountered by genome architecture changes), or exogenous genotoxic factors (ultraviolet radiation, genotoxic agents, etc.). Genoautotomy only causes nonlethal DNA damage, which involves cutting and releasing single-stranded DNA (ssDNA) fragments from double-stranded DNA, and leaving ssDNA gaps in the genome. The process of this DNA damage can help to remove genotoxic agents or DNA lesions from the genome or to overcome the obstacles encountered by genome architecture changes during cell cycle progression, and thus prevents the whole genome from lethal damage. The ssDNA gaps in the genome will be precisely repaired later. The fate of released ssDNA fragments, by-products of genoautotomy, could vary widely depending on their size, quantity, and sequence. Some might be completely degraded by ssDNA specific nuclease upon releasing from the genome; others might survive the nuclease by binding to ssDNA-binding proteins [7]; certain short free-form ssDNA fragments including those derived from partial nuclease digestion or programmed splicing of large ssDNA fragments could downregulate gene expression by binding to complementary ssDNA region of transcription site to block transcription or by serving as intrinsic antisense oligonucleotides to block translation. It is also possible that some small ssDNA fragments might serve as intrinsic mutation-contained primers involved in mutagenesis or natural directed mutagenesis and thus play an important role in eukaryotic evolution and drug resistance development [8].

The genoautotomy proposed in this paper is supported by 4 types of research evidence. First, nucleotide excision repair (NER) is a ssDNA damage repair mechanism, which involves ssDNA lesion excision, releasing short or long lesion-contained ssDNA fragments [9]. This excision can be considered as a type of genome "self-injury" which contributes to the maintenance of genome integrity, therefore, NER belongs to genoautotomy. Second, anticancer agents, adriamycin and daunomycin, can induce ssDNA regions in nuclear DNA [10], and many long ssDNA molecules were found in the Chinese hamster cells treated by anticancer agent, hydroxyurea, which is confirmed by electron microscopy [11]. The ssDNA regions and ssDNA molecules produced by anticancer agents can be considered as the results of cellular defence response to the exogenous genotoxic agents as ssDNA gaps and released ssDNA fragments can help to remove the agents from the genome. Furthermore, the released ssDNA fragments could also slow down the progress of cell cycle progression, which is necessary for cell to go through tough times. Therefore, anticancer agent induced ssDNA regions and released ssDNA fragments could also be considered as a kind of genoautotomy. Third, one study showed that DNA isolated from Chinese hamster ovary cells synchronized in $G_1$ phase was found to contain ssDNA regions, separated by a distance of 100 μm. The author emphasized that the ssDNA regions in the genome of $G_1$ cells were demonstrated not to be the result of a low level of DNA replication nor to be an artifact of the isolation procedure [12]. It is reasonable to think that the ssDNA regions might be caused by genoautotomy that was triggered by unknown endogenous genotoxic factors. Fourth, our experimental results showed that *Plasmodium falciparum* produced certain cell-cycle-associated amplified genomic-DNA fragments (CAGFs) during cell cycle progression, and DNA intercalating agent, chloroquine, could induce the production of CAGFs [13]. Since CAGFs are easily degraded, they are thought to be ssDNA fragments produced by endonuclease dependent transcript cutout, which is a kind of genome "self-injury" [3]. Our research findings indicate that both exogenous genotoxic factor, such as chloroquine, and endogenous genotoxic factor (unknown) involve in triggering genoautotomy in *P. falciparum*.

The main function of genoautotomy is to help the cells survive severe genotoxic stress. Therefore, small number of ssDNA gaps and released ssDNA fragments derived from genoautotomy will not cause cell death. However, severe genoautotomy might elicit cell growth arrest or cell death because too many ssDNA gaps might not be completely repaired on time, and unpaired ones might be further damaged into double strand breaks, which can trigger apoptosis [14]. Besides, too many ssDNA fragments might greatly downregulate genes that control cell cycle progression, and obstruct genome architecture changes, which could lead to cell growth arrest or cell death. One study has reported that introduction of ssDNA fragments into eukaryotic cells can induce DNA damage as well as apoptosis signals [15].

Taken together, genoautotomy (genome "self-injury") is the process of programmed DNA damage in eu-

...

karyotic cells, which prevents the whole genome from lethal damage, and thus should be deemed as a cellular defence response to genotoxic stress. However, genoautotomy is a double-edged sword, its initial function is to help the whole genome survive severe genotoxic stress, but excess genoautotomy could make the genome more vulnerable to further attacks, or even cause cell death. Some anticancer agents and chloroquine might exert their toxic effects by triggering excess genoautotomy.

## 3. Idea for Further Investigation

Genoautotomy proposed in this paper is a hypothesis. To validate this hypothesis, further investigation should be done. A simple experiment to prove the existence of genoautotomy in eukaryotic cells can be carried out in this way: a suitable cell line is treated with sub-lethal dose of genotoxic agent and then to check if ssDNA gaps in the genome and ssDNA fragments or molecules in the nucleus are increased. If both ssDNA gaps and ssDNA fragments are increased, then check if they are greatly reduced or completely disappear after withdrawal of the genotoxic agent. A wide range of techniques including PCR, immunological techniques and electron microscopy can be used in this experiment.

Once existence of genoautotomy in eukaryotic cells is confirmed, the pathway underling genoautotomy should be comprehensively investigated. Presumably, genoautotomy might share some pathways with apoptosis and autophagy. A detailed investigation of the interplay among these three players might contribute to understanding of cellular defence functions and designing of efficient cancer therapy.

Majority of ssDNA fragments produced by genoautotomy or derived from other origins (DNA replication, DNA repair and transcription site) will be degraded by ssDNA specific nuclease as too many ssDNA fragments inside the nucleus are harmful to the maintenance of genome integrity. However, since ssDNA binding proteins inside the nucleus can protect ssDNA fragments from nuclease degradation, the normal eukaryotic cells might maintain a low-level concentration of ssDNA fragments inside the nucleus as an adaptive trait. Under certain conditions, the ssDNA fragments could be released as a free form to exert their effects on cell cycle progression. Further study in this area might be able to identify a new mechanism of epigenetic regulation of gene expression in eukaryotic cells.

## 4. Implications of Genoautotomy

Like apoptosis ("self-killing") and autophagy ("self-eating"), genoautotomy (genome "self-injury") might be another important cellular defence function in eukaryotic cells. Uncovering the pathway underlying genoautotomy will enrich our knowledge of cellular functions and thus fundamentally contributes to life science research.

## 5. Conclusion

Based on some research findings, genoautotomy (genome "self-injury"), the process of programmed DNA damage, has been proposed in this paper, which could be deemed as a eukaryotic cellular defence response to genotoxic stress caused by endogenous or exogenous genotoxic factors. Further exploration in this area might contribute to understanding of genome protection, gene regulation, mutagenesis and the mechanism of action of some anticancer agents.

## Conflict of Interest

The author declares that there is no conflict of interest regarding the publication of this paper.

## References

[1] Maiuri, M.C., Zalckvar, E., Kimchi, A. and Kroemer, G. (2007) Self-Eating and Self-Killing: Crosstalk between Autophagy and Apoptosis. *Nature Reviews Molecular Cell Biology*, **8**, 741-752. http://dx.doi.org/10.1038/nrm2239

[2] Rubinstein, A.D. and Kimchi, A. (2012) Life in the Balance—A Mechanistic View of the Crosstalk between Autophagy and Apoptosis. *Journal of Cell Science*, **125**, 5259-5268. http://dx.doi.org/10.1242/jcs.115865

[3] Li, G.D. (2016) A Possible Mechanism of DNA to DNA Transcription in Eukaryotic Cells: Endonuclease Dependent Transcript Cutout. *Open Access Library Journal*, **3**, e2758. http://dx.doi.org/10.4236/oalib.1102758

[4] Li, G.D. (2016) DNA to DNA Transcription Might Exist in Eukaryotic Cells. *Open Access Library Journal*, **3**, e2665.

http://dx.doi.org/10.4236/oalib.1102665

[5] Clause, A.R. and Capaldi, E.A. (2006) Caudal Autotomy and Regeneration in Lizards. *Journal of Experimental Zoology. Part A*, *Comparative Experimental Biology*, **305**, 965-973. http://dx.doi.org/10.1002/jez.a.346

[6] Fleming, P.A., Muller, D. and Bateman, P.W. (2007) Leave It All behind: A Taxonomic Perspective of Autotomy in Invertebrates. *Biological Reviews of the Cambridge Philosophical Society*, **82**, 481-510. http://dx.doi.org/10.1111/j.1469-185X.2007.00020.x

[7] Dickey, T.H., Altschuler, S.E. and Wuttke, D.S. (2013) Single-Stranded DNA-Binding Proteins: Multiple Domains for Multiple Functions. *Structure*, **21**, 1074-1084. http://dx.doi.org/10.1016/j.str.2013.05.013

[8] Li, G.D. (2016) "Natural Site-Directed Mutagenesis" Might Exist in Eukaryotic Cells. *Open Access Library Journal*, **3**, e2595. http://dx.doi.org/10.4236/oalib.1102595

[9] Schärer, O.D. (2013) Nucleotide Excision Repair in Eukaryotes. *Cold Spring Harbor Perspectives in Biology*, **5**, a012609. http://dx.doi.org/10.1101/cshperspect.a012609

[10] Center, M.S. (1979) Induction of Single-Strand Regions in Nuclear DNA by Adriamycin. *Biochemical and Biophysical Research Communications*, **89**, 1231-1238. http://dx.doi.org/10.1016/0006-291X(79)92140-5

[11] Carnevali, F. and Filetici, P. (1981) Single-Stranded Molecules in DNA Preparations from Cultured Mammalian Cells at Different Moments of Cell Cycle. *Chromosoma*, **82**, 377-384. http://dx.doi.org/10.1007/BF00285763

[12] Henson, P. (1978) The Presence of Single-Stranded Regions in Mammalian DNA. *Journal of Molecular Biology*, **119**, 487-506. http://dx.doi.org/10.1016/0022-2836(78)90198-5

[13] Li, G.D. (2016) Certain Amplified Genomic-DNA Fragments (AGFs) May Be Involved in Cell Cycle Progression and Chloroquine Is Found to Induce the Production of Cell-Cycle-Associated AGFs (CAGFs) in *Plasmodium falciparum*. *Open Access Library Journal*, **3**, e2447. http://dx.doi.org/10.4236/oalib.1102447

[14] Kaina, B. (2003) DNA Damage-Triggered Apoptosis: Critical Role of DNA Repair, Double-Strand Breaks, Cell Proliferation and Signaling. *Biochemical Pharmacology*, **66**, 1547-1554. http://dx.doi.org/10.1016/S0006-2952(03)00510-0

[15] Nur-E-Kamal, A., Li, T.K., Zhang, A., Qi, H., Hars, E.S. and Liu, L.F. (2003) Single-Stranded DNA Induces Ataxia Telangiectasia Mutant (ATM)/p53 Dependent DNA Damage and Apoptotic Signals. *Journal of Biological Chemistry*, **278**, 12475-12481. http://dx.doi.org/10.1074/jbc.M212915200